\documentclass[reprint,showpacs,preprintnumbers,amsmath,a4paper,amssymb,xcolor=dvipsnames]{revtex4-1}
\pdfoutput=1

\usepackage[pdftex]{graphicx}
\usepackage{graphicx}
\usepackage{dcolumn}
\usepackage{bm}
\usepackage{datetime}
\usepackage{float}
\usepackage[pdftex,colorlinks=true,urlcolor=black,linkcolor=black]{hyperref} 
\usepackage{microtype}
\usepackage{epstopdf} 
\usepackage{units}
\usepackage{amsmath}
\usepackage{amssymb}
\usepackage{xcolor}

\begin{document}

\title{Two Fermions in a double well:\\Exploring a fundamental building block of the Hubbard model}


	\title{Two Fermions in a Double Well:\\Exploring a Fundamental Building Block of the Hubbard Model} 

\author{Simon Murmann}
	\thanks{These authors contributed equally to this work.}
\author{Andrea Bergschneider}
	\thanks{These authors contributed equally to this work.}
\author{Vincent M. Klinkhamer}
\author{Gerhard Z\"urn}
\author{Thomas Lompe}
	\altaffiliation[Present address: ]{MIT-Harvard Center for Ultracold Atoms, Research Laboratory of Electronics, and Department of Physics, Massachusetts Institute of Technology, Cambridge, Massachusetts 02139, USA}
\author{Selim Jochim}
	\affiliation{Physikalisches Institut der Universit\"at Heidelberg, Im Neuenheimer Feld 226, 69120 Heidelberg, Germany}

\begin{abstract}
We have prepared two ultracold fermionic atoms in an isolated double-well potential and obtained full control over the quantum state of this system. In particular, we can independently control the interaction strength between the particles, their tunneling rate between the wells and the tilt of the potential. By introducing repulsive (attractive) interparticle interactions we have realized the two-particle analog of a Mott-insulating (charge-density-wave) state. We have also spectroscopically observed how second-order tunneling affects the energy of the system. This work realizes the first step of a bottom-up approach to deterministically create a single-site addressable realization of a ground-state Fermi-Hubbard system.

\end{abstract}

\maketitle

In the presence of strong correlations, the understanding of quantum many-body systems can be exceedingly difficult. One way to simplify the description of such systems is to use a discrete model where the motion of the particles is restricted to hopping between the sites of a lattice. The paradigmatic example for this approach is the Hubbard model, which reduces the physics of a quantum many-body system to tunneling of particles between adjacent sites and interactions between particles occupying the same site. While this model captures essential properties of electrons in a crystalline solid and provides a microscopic explanation for the existence of Mott-insulating and antiferromagnetic phases, many questions about this Hamiltonian --- such as whether it can explain d-wave superfluidity --- are still unanswered \cite{Anderson1987}.

A promising approach to answer these questions is to use ultracold atoms trapped in periodic potentials as quantum simulators of the Hubbard model \cite{Jaksch1998, Jordens2008, Schneider2008, Bakr2010, Sherson2010, Esslinger2010, Pertot2014}. Such experiments have been performed both in large- and small-scale systems. Degenerate gases loaded into optical lattices have been used to observe the transition to the bosonic \cite{Greiner2002, Stoferle2004} and fermionic Mott insulator \cite{Jordens2008, Schneider2008}. The first observation of second-order tunneling was achieved in a small-scale system by studying the tunneling dynamics of bosonic atoms in an array of separated double wells \cite{Folling2007, Trotzky2008}. In a recent experiment, these two regimes have been connected by splitting a fermionic Mott insulator into individual double wells. In this way, the strength of the antiferromagnetic correlations in the many-body system could be determined by measuring the fraction of double wells with two atoms in the spin-singlet configuration \cite{Greif2013}. But despite the observation of antiferromagnetic correlations\cite{Greif2013, Hart2014} current experiments using fermionic atoms have so far failed to reach temperatures below the critical temperature of spin ordering \cite{McKay2011, Jordens2010}.

Recently, new experimental techniques have been developed which allow for the deterministic preparation of few-particle systems in the ground state of a single potential well \cite{Serwane2011, Kaufman2012, Kaufman2014}. This makes it feasible to use ultracold atoms to study many-body physics in a bottom-up approach, i.e. to start from the fundamental building block of the system and watch how many-body effects emerge as one gradually increases the system's size \cite{Wenz2013}. Here we report on the realization of the fundamental building block of the Fermi-Hubbard model at half filling, which consists of one $\left|\uparrow\right\rangle$ and one $\left|\downarrow\right\rangle$ particle in a spin-singlet configuration in a double-well potential.

\begin{figure*}
        \centering
        \includegraphics[width=0.8\textwidth]{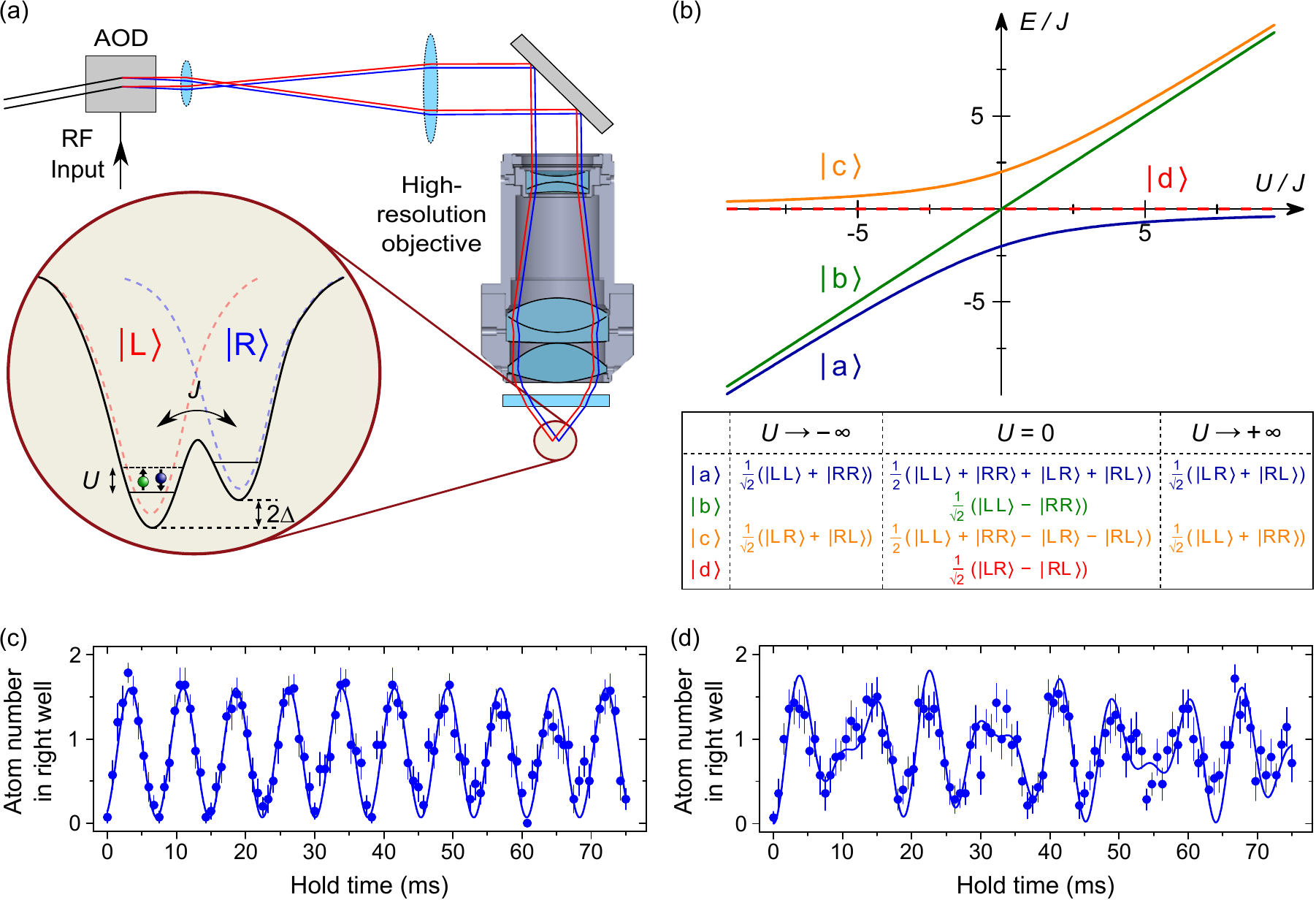}
        \caption{Experimental realization and eigenenergies of the two-site Hubbard model. (a) Experimental setup: The double-well potential is created by focusing two laser beams with a high-resolution objective. By independently controlling the intensity and position of the two laser beams with an acousto-optic deflector (AOD) we can tune the tunnel coupling $J$ and the tilt $\Delta$ between the two wells. (b) Energies of the four lowest two-particle eigenstates in a symmetric double-well potential as a function of the on-site interaction energy $U$. (c, d) Tunneling of two particles in the double well. The data show the time evolution of the particle number in the right well after initializing the system with both particles in the left well and abruptly switching on the tunnel coupling $J$ between the two sites. For $U\approx 0$ and $\Delta \approx 0$ (c), we can extract the value of the tunnel coupling by fitting the data with a damped sine wave. For intermediate interaction strength ($U \approx J$) (d), we observe correlated tunneling of the two particles, which shows good agreement with the prediction from the Hubbard model (solid line). The error bars denote the $1\sigma$ statistical uncertainty. }
        \label{fig:first_graph}
\end{figure*}

\begin{figure}
        \centering
                \includegraphics[width=0.38\textwidth]{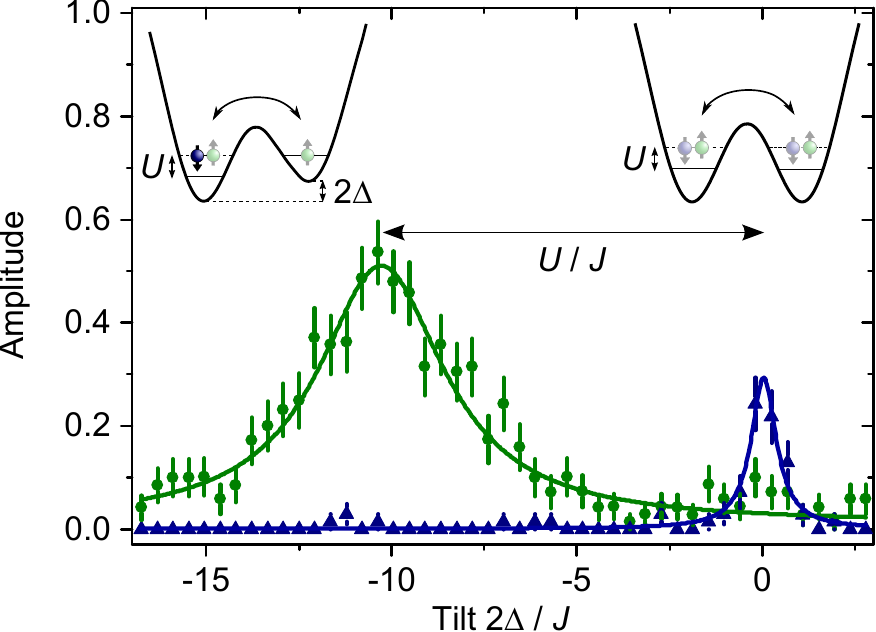}
        \caption{Single-particle and pair tunneling as a function of the tilt $\Delta$ at an interaction energy of $U/J = \unit{10.05 \pm 0.19}$. The data points show the time-averaged probability of finding a single particle (green circles) or a pair of particles (blue triangles) in the right well after initializing the system with two atoms in the left well and switching on the tunneling. Pair tunneling is resonant in a symmetric double well, while conditional single-particle tunneling occurs for a tilt of $- 2 \Delta = U$. The error bars denote the $1\sigma$ statistical uncertainty.}
        \label{fig:second_graph}
\end{figure}

By starting from this two-site realization of the Hubbard model, we can test our building block in a regime where the model can still be easily solved. In the Hubbard regime, the spatial wave function $\left|\Psi\right\rangle$ of this two-particle system can be written in the basis $\{ \, \left|LL\right\rangle , \left|LR\right\rangle , \left|RL\right\rangle , \left|RR\right\rangle \, \}$. These basis states are all possible combinations of the localized single-particle states $\left|L\right\rangle$ and $\left|R\right\rangle$ of one particle in the ground state of either the left or the right well. In this basis, the spatial part of the Hamiltonian is
\begin{equation}
  \begin{aligned}
		H = \begin{pmatrix} {U + 2\Delta}&{-J}&{-J}&{0}\\{-J}&{0}&{0}&{-J}\\{-J}&{0}&{0}&{-J}\\{0}&{-J}&{-J}&{U - 2\Delta} \end{pmatrix} ,
  \end{aligned}
	\label{formula2}
\end{equation}
with the tunneling matrix element $J$, the on-site interaction energy $U$ and the energy tilt $2 \Delta$ between the wells. Diagonalizing this Hamiltonian leads to three eigenstates ($\left|a\right\rangle$, $\left|b\right\rangle$ and $\left|c\right\rangle$) which are symmetric and one eigenstate ($\left|d\right\rangle$) which is antisymmetric with respect to particle exchange (Fig\,1b).

To prepare our double-well system, we start with two $^6$Li atoms in different hyperfine states in the motional ground state of a single optical microtrap \cite{Serwane2011} (Sec. I of the Supplemental Material \cite{SOM}). We then slowly ramp on a second potential well and thereby deform our single trap into a double-well potential. During this process, we keep the coupling between the wells negligible and thus initialize the system in state ($\left|LL\right\rangle$) where both atoms reside in the ground state of the left well. This state is the starting point for all our measurements. We can prepare it with a fidelity of more than $\unit[90]{\%}$. The predominant error is that there is only one atom in the trapping potential, while the probability to start with three atoms is $\lesssim \unit[1]{\%}$.
\begin{figure*}
        \centering
               \includegraphics[width=0.8\textwidth]{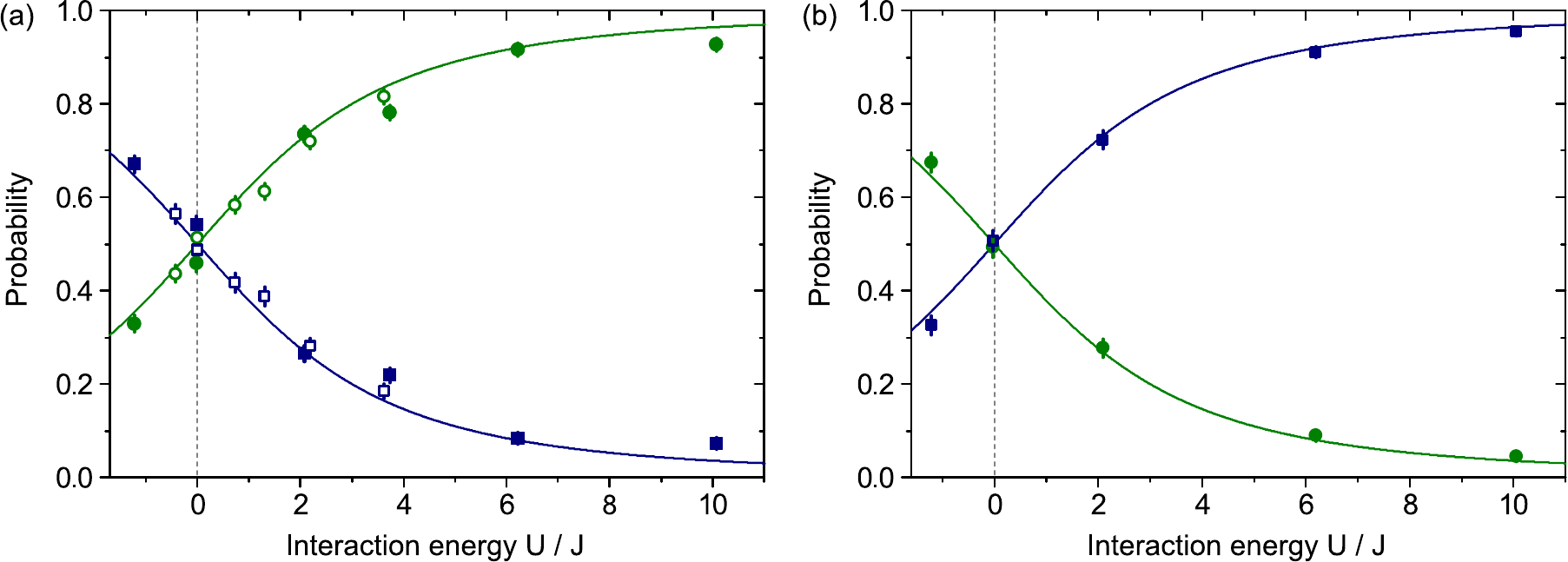}
        \caption{Occupation statistics as a function of interaction strength. The relative probabilities of measuring both particles in the same well ($P_2$, blue squares) or in different wells ($P_1$, green circles) are shown as a function of the on-site interaction energy $U$. Open (filled) symbols indicate a tunnel coupling of $J/h \simeq \unit[142]{Hz}$ ($J/h \simeq \unit[67]{Hz}$), the solid lines show the prediction of the Hubbard model. (a) For the ground state $\left|a\right\rangle$, double occupancy is suppressed for increasing repulsive interactions. This indicates the crossover from a metallic to a Mott-insulating regime. For attractive interactions, double occupancy is enhanced, which we interpret as the onset of a charge-density-wave regime. (b) For the excited state $\left|c\right\rangle$, we observe the crossover to the charge-density-wave regime for strong repulsive interactions. For both measurements, the data have been corrected for the effect of the finite fidelities of preparation and detection (Sec. VII and Fig. S5 of \cite{SOM}). The error bars denote the $1\sigma$ statistical uncertainty.}
        \label{fig:third_graph}
\end{figure*}

\begin{figure}
        \centering
                \includegraphics[width=0.38\textwidth]{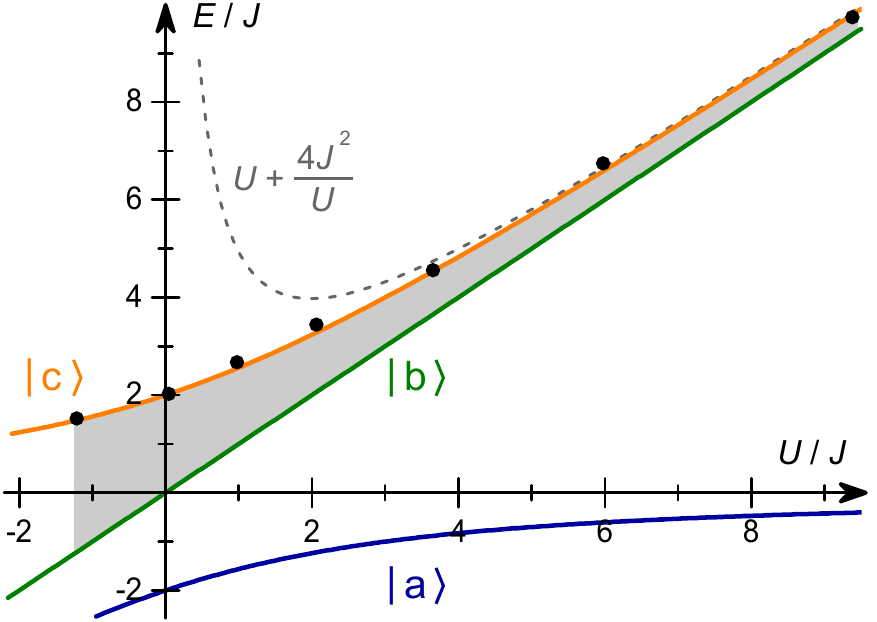}
        \caption{Transition from first-order to second-order tunneling. To directly measure the influence of the tunnel coupling on the energy of the system, we determine the energy difference $E_\text{bc} = E_\text{c} - E_\text{b}$ (shaded region) between state $\left|c\right\rangle$ (orange) and state $\left|b\right\rangle$ (green) using trap modulation spectroscopy. To compare our results to the eigenenergies of the Hubbard model (Fig. 1b), we plot the sum of $E_\text{bc}$ and the energy of state $\left|b\right\rangle$ ($E_\text{b}\,=\,U$). The error bars for the $1\sigma$ statistical uncertainties (Sec. VIII of \cite{SOM}) are smaller than the symbols of the data points. For $U=0$, we observe that $E_{bc} \approx 2J$, which is consistent with single-particle tunneling. For increasing repulsion, first-order tunneling is suppressed and $E_{bc}$ converges to the superexchange energy $4J^2/U$ (see dashed line).}
        \label{fig:fourth_graph}
\end{figure}

Since the spatial wave function of the initial state is symmetric with respect to particle exchange, the spin wave function of the two fermions is in a singlet configuration. In the experiments presented here, we do not couple the position and the spin of the particles and therefore restrict the spatial wave function of the system to the symmetric eigenstates $\left|a\right\rangle$, $\left|b\right\rangle$ and $\left|c\right\rangle$.

In a first set of experiments, we study the tunneling dynamics of the two particles in our double-well potential to characterize the Hubbard parameters $J$, $U$, and $\Delta$ of our system. To do this, we initialize the system in state $\left|LL\right\rangle$ and abruptly reduce the height of the potential barrier, which allows the atoms to tunnel between the wells. To observe the resulting dynamics, we let the system evolve for different durations and then freeze the spatial distribution of the atoms by quickly increasing the barrier height. We then count the number of atoms in one of the wells by recapturing them into a magneto-optical trap and measuring their fluorescence \footnote{Note that this technique allows us to measure the atom number without parity projection.} \cite{Serwane2011} (Sec. III of \cite{SOM}).

For a noninteracting system and a small tilt of the double well ($|\Delta| \lesssim J$), we observe long-lived tunneling oscillations whose frequencies we can set by tuning the barrier height (Sec. V.A, Fig. S1 and Fig. S2(a) of \cite{SOM}). As an example, a typical oscillation for $\Delta \simeq 0$ with a frequency of $2J/h \simeq \unit[134]{Hz}$ is shown in figure\,1c. To calibrate $\Delta$, we measure the oscillation frequency for different tilts and find good agreement with the effective coupling strength $J_\text{eff} = \sqrt{J^2+\Delta^2}$ of a two-level system (Sec. V.B and Fig. S2(b) of \cite{SOM}).

For an interacting system, the interaction energy creates an effective detuning for the tunneling of a single particle. In the limit of strong interactions and a symmetric double well ($U \gg J$ and $\Delta=0$), the atoms can therefore only tunnel as pairs \cite{Winkler2006}. However, we can restore single-particle tunneling by setting a tilt that compensates for the interaction shift \cite{Folling2007}. This allows us to calibrate the on-site interaction energy $U$ by measuring the strength of single-particle tunneling as a function of the tilt $\Delta$ for different interaction strengths (Fig.\,2). We find good agreement with a calculation of the interaction energy of two particles in the single well \cite{Idziaszek2006} (Sec. V.C and Fig. S3 of \cite{SOM}).

Using the parameters determined from these measurements, the two-site Hubbard Hamiltonian (Eq. 1) fully describes the oscillations of two interacting particles in our double well (solid line in Fig.\,1d).

To use our system as a fundamental building block of the Hubbard model, we must be able to prepare the particles in the ground state of the symmetric double well. To achieve this, we first initialize the system with both atoms in the left well and a tilt $\Delta \ll 0$. Then, we adiabatically change the tilt to bring the system into the ground state of the symmetric double well \cite{Petta2004} (Sec. VI and Fig. S4 of \cite{SOM}) \footnote{We have verified that the probability of exciting atoms to higher trap levels while changing the potential is smaller than $\unit[1]{\%}$.}.

To show that our fundamental building block already contains the physics that is responsible for the formation of ordered phases in a many-body system, we first measure the influence of the interaction energy $U$ on the distribution of the two particles between the wells. We therefore determine the probabilities $P_1=|\langle \Psi|LR \rangle |^2+|\langle \Psi|RL \rangle |^2$ of finding the two particles on different sites (single occupancy) and $P_2=|\langle \Psi|LL \rangle |^2+|\langle \Psi|RR \rangle |^2$ of finding both particles on the same site (double occupancy) by measuring the probabilities of having 0, 1 or 2 atoms in one of the wells (Sec. VII of \cite{SOM}).

In a noninteracting system ($U=0$), the spatial wave function of two particles in the ground state is an equal superposition of the basis states $\left|LL\right\rangle$, $\left|LR\right\rangle$, $\left|RL\right\rangle$, and $\left|RR\right\rangle$. This leads to equal probabilities $P_1$ and $P_2$ which we observe in our measurements (Fig.\,3a).

In a system with strong repulsive on-site interactions, it is energetically unfavorable to have two atoms occupying the same site. The ground state of our system then becomes a two-particle analog to a Mott-insulating state which we observe as a reduction of double occupancy. For attractive interactions we observe an increase in double occupancy that marks the onset of a paired state. We interpret this state as the two-particle limit of the cold gas analog of a charge-density-wave state as described in \cite{Ho2009}. We perform these measurements for two different barrier heights. For the larger barrier height ($J/h \simeq \unit[67]{Hz}$), we find good agreement with the prediction of the Hubbard model (Fig.\,3a). For the smaller barrier height ($J/h \simeq \unit[142]{Hz}$), we observe a small deviation from the theoretical expectations which might indicate that our system is approaching the limits of the Hubbard approximation.

Since in our isolated, two-particle system excited states are stable against relaxation, we can also prepare the system in the highest-energy eigenstate $|c\rangle$ (Fig. S4 of \cite{SOM}). When measuring the occupation statistics of this state as a function of interaction strength, we find that the number of doubly occupied sites is enhanced for repulsive interactions (Fig.\,3b). This allows us to study the charge-density wave regime in a system with repulsive interactions.

In our final measurement, we directly probe the effect of the tunnel coupling on the energy of the system by trap modulation spectroscopy (Sec. VIII of \cite{SOM}). For this, we initialize the system in state $|c\rangle$ and measure the energy difference to state $|b\rangle$ as a function of interaction strength. For $U=0$, the atoms are delocalized over the wells by single-particle tunneling. This leads to a change in the kinetic energy of the system that is proportional to the tunnel coupling $J$ (Fig.\,4). As we increase the interactions, the system enters the insulating regime and first-order tunneling is suppressed. However, second-order tunneling is still possible and we observe a crossover to a new energy scale given by the superexchange energy $4J^2/U$. This energy is directly responsible for the appearance of spin order in the ground state of the Hubbard model \footnote{We cannot use trap modulation spectroscopy to directly measure the superexchange energy between the ground state \unexpanded{$|a\rangle$} and state \unexpanded{$|d\rangle$}, since it does not couple to the spin of the particles.}.

 By combining a series of isolated double wells we can realize a dimerized lattice \cite{Greif2013} where each dimer contains two fermions in a spin-singlet configuration with high fidelity. By adiabatically lowering the barriers between the double wells, we can then prepare a low-entropy state in a homogeneous lattice. Since our systems can be prepared with arbitrary filling factors they are also ideally suited to study the effects of hole doping. Additionally, the tunability of our potential allows us to explore finite-size lattices with arbitrary geometries \cite{Zimmermann2011} and introduce controlled disorder into our system.  Finally, our experiments provide a starting point for scalable quantum computation with neutral atoms \cite{Raizen2009, Anderlini2007, Isenhower2010, Wilk2010a}. \\

\acknowledgments{Supported by the European Research Council starting grant 279697 and the Heidelberg Center for Quantum Dynamics.}
%
%


\newpage
\phantom{}
\newpage
\section*{Supplemental material} 

\setcounter{figure}{0}

\section{Preparation of two fermions in a microtrap}
\label{sec:prep}

We start our experiments by preparing two $^6$Li atoms in the $|F=1/2, m_\text{F}=+1/2\rangle$ and $|F=1/2, m_\text{F}=-1/2\rangle$ hyperfine states -- labeled $\left|\uparrow\right\rangle$ and $\left|\downarrow\right\rangle$ -- in the motional ground state of an optical microtrap. To do this we follow the procedure established in Serwane et al.\cite{Serwane2011}. We first prepare a two-component Fermi gas consisting of $N_{|\uparrow\rangle} \approx N_{|\downarrow\rangle} \approx 2.5\times 10^4$\ atoms in a crossed-beam optical dipole trap at a temperature of $ T \approx 250$\,nK. This dipole trap acts as a reservoir from which we load about 1000 atoms into our microtrap, which is created by focusing a far red-detuned laser beam ($\lambda = 1064$\,nm) with a custom-designed high-resolution objective ($\text{NA}=0.55$)\cite{Serwane2011a}.  We then apply a magnetic field gradient to deform the trapping potential until only a single bound state remains in the microtrap. This spilling technique allows us to prepare systems of one $\left|\uparrow\right\rangle$ and one $\left|\downarrow\right\rangle$ atom in the ground state of the trap with a typical fidelity above $90\%$.

To characterize our microtrap, we have measured its trapping frequencies using modulation spectroscopy. We find trap frequencies of $\omega_\text{ax} \approx 2\pi\times2.5$\,kHz and $\omega_\text{rad} \approx 2\pi\times16.5$\,kHz at an optical power of $P\approx 393\,\mu$W, which is in good agreement with our expectations for a cigar-shaped dipole trap with a waist of $w_0 \approx 1.66\,\mu m$ and an aspect ratio  $\eta\approx 7$.

\section{Creation of the double-well potential}
\label{sec:dwpotential}

To form a double-well potential, we create two microtraps which are partially overlapping. We achieve this by using an acousto-optical deflector (AOD) to create two separated laser beams that are incident upon the objective under slightly different angles, which the objective translates into a spatial separation of the two foci in the focal plane. To control the spacing and the tilt of the resulting double-well potential we tune the relative angle and intensity of the two trapping beams. We do this by changing the frequencies and amplitudes of two RF-signals driving the AOD. To obtain a tunnel coupling which is on the order of 100\,Hz at our typical light powers of $200\,\mu$W, we set the frequencies to values of 32\,MHz and 38\,MHz which corresponds to a spacing of the double well of $d \approx 2\,\mu$m. To tune the strength of the tunnel coupling during our experiments, we change the overall depth of the potential which we control by actively stabilizing the overall light power $P_{\text{tot}}$ of the trapping beams. At these trap parameters, the individual wells have a depth of approximately $h\times\unit[30]{kHz}$ and the height of the potential barrier between them is approximately $h\times\unit[5.4]{kHz}$. To compare our double well to optical lattice experiments, we can define an effective lattice wavelength $\lambda_\text{eff} = 2 d$ and calculate the recoil energy $E_\text{rec} = h^2 / 2 m \lambda_\text{eff}^2 \approx h\times\unit[2.1]{kHz}$, where $h$ is the Planck constant and $m$ is the mass of $^6$Li.

\section{Site-selective atom number detection}
\label{sec:detection}

Since our experiment allows us to independently address the individual wells of our potential, we can perform site-selective atom-number detection by switching off one of the wells and measuring the number of atoms in the remaining well. 

To do this, we first decouple the wells by rapidly increasing the height of the potential barrier within 2\,ms. This ramp is fast compared to the time scale of the tunnel coupling but slow enough to avoid heating of the atoms into higher trap levels. Then we slowly turn off one of the wells while keeping the overall light power constant. During this process, the atoms in this well are adiabatically transferred into excited states of the remaining well. Therefore, we use the spilling technique introduced above to remove all atoms from these highly excited states.

To count the number of atoms in the remaining well, we release the atoms from the microtrap, recapture them in a magneto-optical trap and measure their fluorescence with a CCD camera (Andor iXon DV887) for an exposure time of 1 second \cite{Serwane2011}.

\section{The two-site Hubbard model}
\label{sec:hubbard}

To experimentally realize the Hubbard model, our double-well system has to fulfill the tight-binding approximation and the single-band approximation.

In a lattice potential, the wavefunction of a single particle can be described as a superposition of localized wavefunctions centered on the individual sites of the lattice. In the limit of tight binding, the overlap between wavefunctions located on different sites is small and they can be approximated by the eigenstates of the individual wells. In this case, the tunnel coupling $J$ between adjacent sites is much smaller than the lowest excitation energy on the individual wells. Since in our double well, the tunneling constant $J$ is approximately $h\times\unit[100]{Hz}$ and the axial trap frequency $\omega_\text{ax}$ in each well is about $2\pi\times\unit[1]{kHz}$, the tight-binding approximation is well fulfilled in our system.

In the single-band approximation, one only considers the motion of the particles in the lowest Bloch band, which is composed of the ground states of the individual wells. To fulfill this approximation, all energy scales in the system have to be much smaller than the band gap to the first excited band.  In our double-well system, we initialize both atoms in the ground state of one well (see Sec.\,\ref{sec:prep}) and limit the interaction energy $U$ to values smaller than the lowest excitation energy of the single wells during all our experiments. For all but the largest values of the interaction energy we estimate the wavefunction overlap between two interacting particles localized in a single potential well and the respective non-interacting ground state to be above $\unit[97]{\%}$. For the largest interaction energies of $U\approx h \times \unit[650]{Hz}$ for the dataset with $J = \unit[67]{Hz}$ ($U\approx h \times \unit[400]{Hz}$ for $J = \unit[142]{Hz}$) the overlap decreases to $\approx \unit[92]{\%}$ ($\approx \unit[94]{\%}$).

Using these two approximations, the spatial wavefunction of a single particle in our double-well system can be described as a superposition of the ground-state wavefunctions of a particle in the left $|L\rangle$ or the right $|R\rangle$ well. This results in a two-level system where the states $|L\rangle$ and $|R\rangle$ are coupled with a tunnel coupling $J$.

\section{Calibration of the Hubbard parameters}
\label{sec:calibration}

We calibrate the Hubbard parameters of our double well by observing the tunneling dynamics in our system. To determine $J$ and $\Delta$, we study the tunneling of two non-interacting particles between the wells (Sect.\,\ref{sec:tunneling} and  Sect.\,\ref{sec:tilt}). We then determine the on-site interaction energy $U$ for different strengths of the interparticle interaction (Sect.\,\ref{sec:onsite}).

\subsection{Tunneling oscillations}
\label{sec:tunneling}

\begin{figure*}
	\centering
	\includegraphics[width=0.7\textwidth]{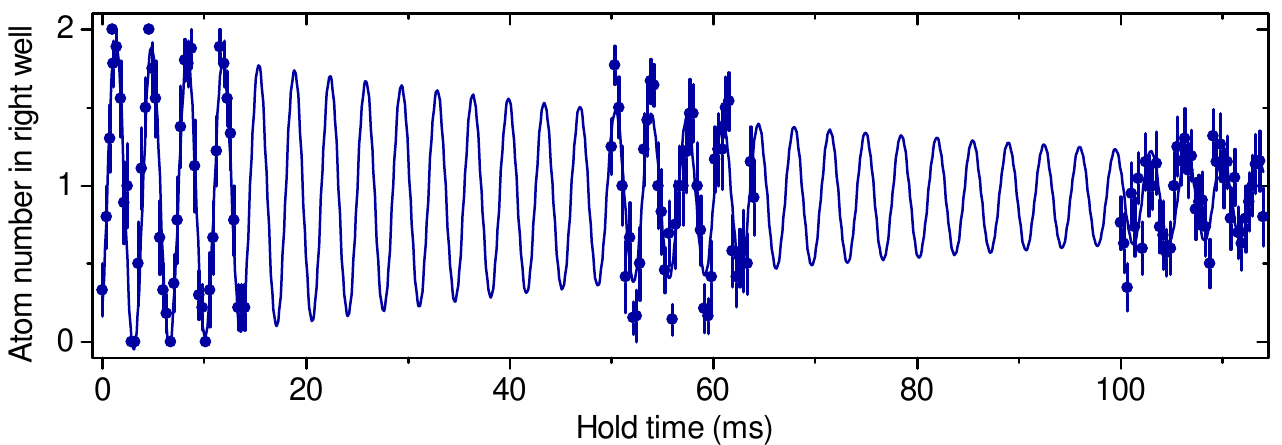}
	\caption{Tunneling of two non-interacting particles in a symmetric double-well potential. Time evolution of the atom number in the right well for a system with two particles in the left well at \mbox{$t=0$}. Each data point is the average of 15 measurements. The solid line shows a damped sinusoidal fit to the data from which we determine the tunnel coupling $J$. The error bars denote the $1\sigma$ statistical uncertainties.}
	\label{fig:som_tunneling_long}
\end{figure*}

To prepare the initial state for our tunneling experiments, we start from two non-interacting atoms in the ground state of a single microtrap (see Sec.\,\ref{sec:prep}). Then we slowly ramp on a second trapping beam (see Sec.\,\ref{sec:dwpotential}) and thereby change the shape of the potential from a single Gaussian well to a double-well potential. During this ramp, we linearly increase the overall power $P_\text{tot}$ to keep the depth of the initial well constant. At any time in this process, $P_\text{tot}$ is large enough to prevent the atoms from tunneling between the wells.

To start the tunneling dynamics, we lower the height of the barrier between the two wells within $2$\,ms. Since this ramp is fast compared to the time scale of the tunnel coupling we can treat it as a sudden switch-on of the tunneling. Yet, this ramp is slow enough to avoid exciting the atoms into higher bands. We then let the system evolve for different hold times, ramp up the barrier to decouple the wells and finally measure the number of atoms in one of the wells, using the detection scheme described above. 

As an example, the tunneling oscillation of two non-interacting atoms in a symmetric double-well at an overall power of $P_\text{tot} = \unit[131]{\mu W}$ is shown in Fig.\,1. To determine the value of the tunnel coupling $J$ we fit the observed oscillation with a damped sine wave and find $J/h=(142.9\pm0.1)$\,Hz and a damping time of $(83\pm9)$\,ms. We believe that the main cause for the observed damping is dephasing due to long-term drifts of the tilt of the double-well potential. 

To calibrate $J$ for different barrier heights, we perform these measurements for different values of the overall light power $P_\text{tot}$ (Fig.\,2a). For our experiments, we use light powers of $P_\text{tot}=131\,\mu$W and $P_\text{tot}=186\,\mu$W which result in tunnel couplings of  $J/h=(142.0\pm0.5)$\,Hz and $J/h=(67.3\pm0.5)$\,Hz. 

\begin{figure*}
	\centering
	\includegraphics[width=0.47\textwidth]{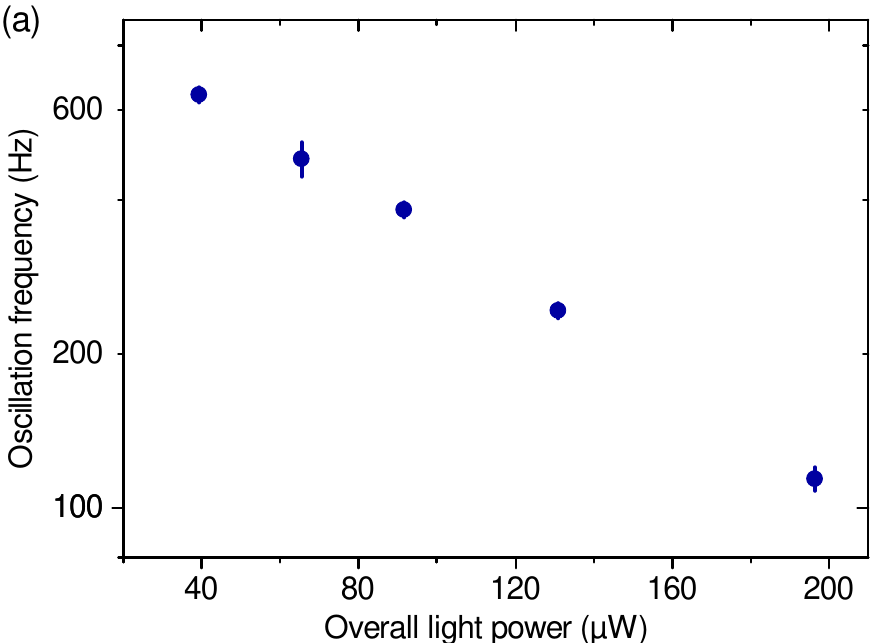}%
	\hspace{3mm}
	\includegraphics[width=0.47\textwidth]{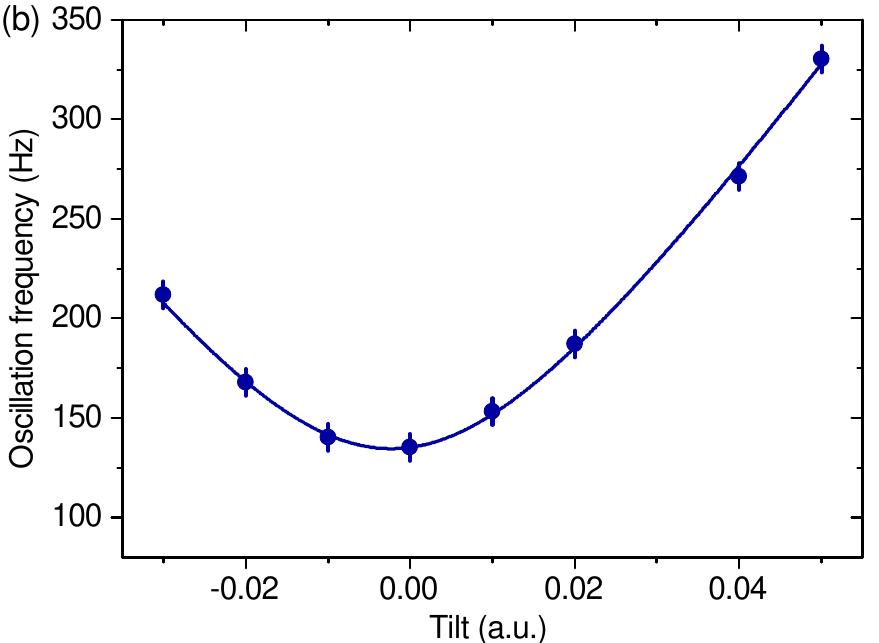}
	\caption{Oscillation frequencies for different parameters of the double well.  (a) Oscillation frequency of two non-interacting particles in a symmetric double well \mbox{($\Delta\,\approx\,0$)} as a function of the overall light power in the trap beams. A decrease of the total light power lowers the barrier height and thereby leads to a faster tunneling of the atoms between the wells. (b)  Oscillation frequency as a function of the relative RF power of the two signals that drive the AOD. The solid line shows a fit with the effective tunnel coupling $2\,J_{\text{eff}}=2\sqrt{ J^2+\Delta^2}$ expected for a two-level system. From the fit, we extract the relative RF power where the double well is balanced. The error bars denote the $1\sigma$ statistical uncertainties.}
	\label{fig:som_first_graph}
\end{figure*}

\subsection{Calibration of the tilt}
\label{sec:tilt}

Tunneling between two states that are separated by a potential barrier is only possible if their energy difference $\Delta$ is not much larger than their tunnel coupling $J$. To achieve this for our double-well system, we control the relative depth of the two wells by changing the relative power of the RF signals driving the AOD (see Sec.\,\ref{sec:dwpotential}).

To calibrate the tilt $\Delta$, we measure the oscillation frequency for various values of the relative RF power (Fig.\,2b)\footnote{Although the AOD can in principle have a nonlinear characteristic, the relative change in RF power in our experiment is so small, that the depths of the wells can be approximated to scale linearly with the RF power.}. We fit the measured frequencies with the effective coupling $2\,J_{\text{eff}}=2\sqrt{ J^2+\Delta^2}$ expected for a two-level system. We find good agreement to the data and therefore conclude that our non-interacting system can be described by the two-site Hubbard model (see Sec.\,\ref{sec:hubbard}). Using this fit, we calibrate the tilt $\Delta$ in units of the tunnel coupling $J$ and extract the relative RF power where the double well is balanced ($\Delta=0$). 

To observe coherent tunneling in our double-well system, we have to stabilize the relative depth of the two wells with a precision better than $J$. For our typical trap depth of $\unit[30]{kHz}$ and tunnel couplings of $J/h \approx \unit[100]{Hz}$ (at an overall light power of $P_\text{tot} \approx \unit[200]{\mu W}$), this corresponds to a relative stability of the intensities of the two beams that is on the order of $10^{-4}$.

However, switching on the second well leads to heating of the AOD which results in a slow drift of $\Delta$ on the order of $h\times\unit[300]{Hz}$ over several seconds. Since these drifts are the same in every experimental cycle, we can compensate them by applying an exponential ramp to the relative RF power. Using this technique, our stability is currently limited by slow drifts of the tilt which are on the order of $\unit[10]{Hz}$ to $\unit[20]{Hz}$ per day.

\subsection{Calibration of the on-site interaction}
\label{sec:onsite}

To control the on-site interaction energy $U$ between two particles that occupy the same well, we change the s-wave scattering length $a$ between the atoms with a magnetic Feshbach resonance. To calibrate the interaction energy as a function of the magnetic offset field, we study the tunneling of two interacting particles.

For a system with two particles in one well, the energy is shifted with respect to the non-interacting system by the interaction energy $U$. If the two particles tunnel together, this interaction energy is conserved and consequently pair tunneling is resonant in a symmetric double well at all interaction strengths. However, if a single particle tunnels to the other well the particles can no longer interact and hence single-particle tunneling is effectively detuned by the interaction energy $U$. We can compensate this effective detuning by a tilt of $2\Delta = -U$ between the two wells\cite{Folling2007}. This allows us to determine the on-site interaction energy by measuring the value of the tilt at which single-particle tunneling is restored.

To perform this calibration, we measure the amplitudes of single-particle and pair tunneling as a function of tilt. To do this we initialize the system with two interacting atoms in the left well, let them tunnel between the wells for different hold times and measure the number of atoms in the right well. From these measurements, we extract the probabilities of finding 1 or 2 atoms in the right well, average over all hold times and plot them as a function of the tilt (Fig.\,2 of main text). We observe that the tilt at which single-particle tunneling is resonant shifts with the interaction strength whereas the pair-tunneling resonance remains at $\Delta=0$. By fitting Lorentzians to the resonance peaks and measuring their separation, we can therefore determine the value of the on-site interaction $U$ for different magnetic fields (Fig.\,3a).

\begin{figure*}
	\centering
	\includegraphics[width=0.47\textwidth]{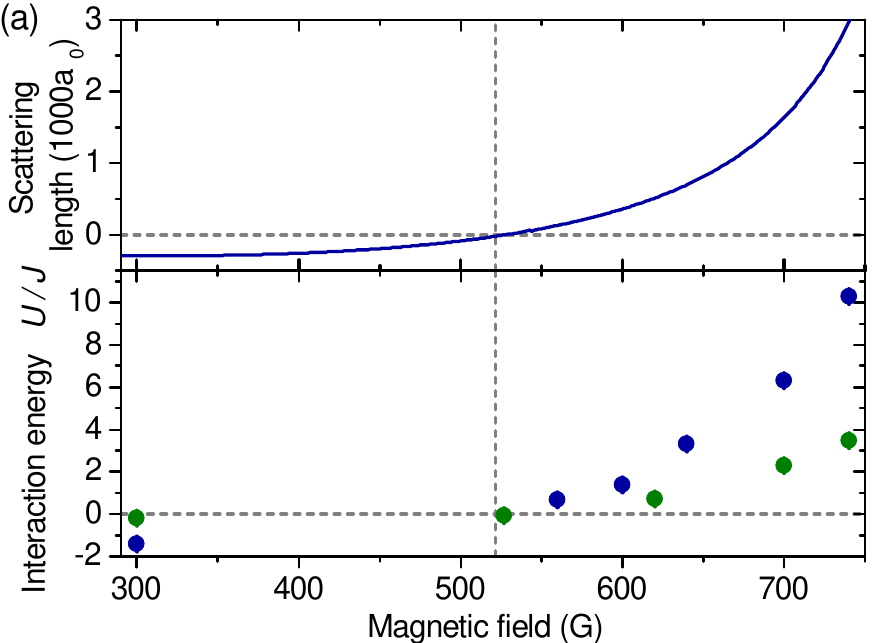}%
\hspace{3mm}
	\includegraphics[width=0.47\textwidth]{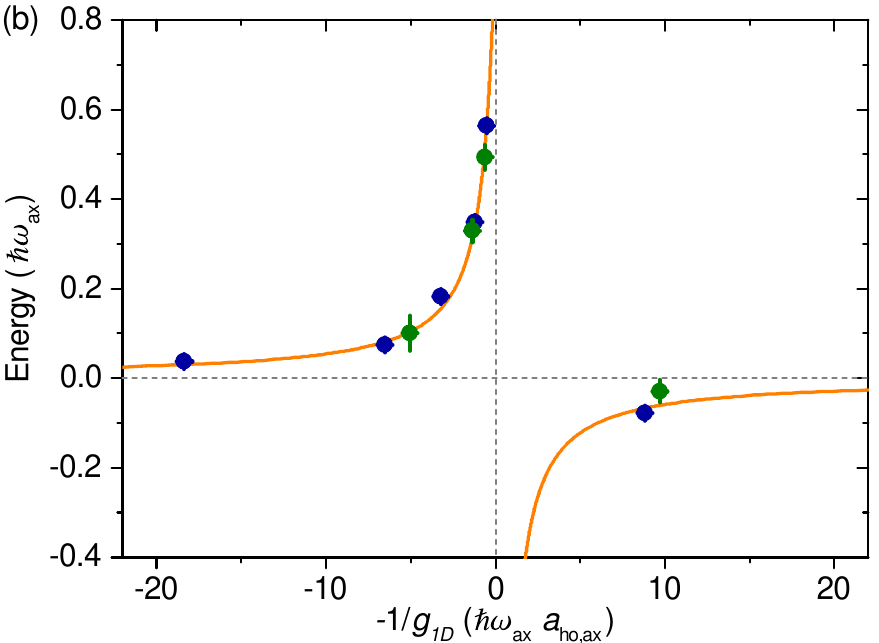}
	\caption{Measurement of the on-site interaction energy $U$ for two different tunnel couplings. (a)  Interaction energy $U$ in units of the tunnel coupling $J$ as a function of the magnetic offset field. The green (blue) dots correspond to a tunnel coupling of $J=h\times(142.0\pm0.5)$\,Hz ($J=h\times(67.3\pm0.5)$\,Hz). (b) Interaction energy $U$ in units of the harmonic oscillator energy $\hbar\omega_\text{ax}$ plotted as a function of the one-dimensional interaction strength $g_\text {1D}$ \cite{Olshanii1998} where $ a_\text{ho,ax} = \sqrt{2 \hbar/(m \omega_\text{ax})}$ denotes the harmonic oscillator length in axial direction and $m$ is the mass of a $^6$Li atom. The data shows good agreement with a calculations of the interaction energy of two particles in a single quasi-1D potential well (solid line)\cite{Idziaszek2006}. The error bars denote the $1\sigma$ statistical uncertainties.}
	\label{fig:som_busch_graph}
\end{figure*}


Since the value of the on-site interaction energy depends on the depth of the confining potential, we perform the calibration of $U$ for the two potential depths (see Sec.\,\ref{sec:onsite}) we use in our experiments. To check whether the measured interaction energies are affected by off-site interactions we compare our results to a calculation of the interaction energy of two particles in a cigar-shaped potential (Fig.\,3b)\cite{Idziaszek2006, Zurn2012}. Within the errors we find good agreement with our data and therefore conclude that the effects of off-site interactions in our experiments are small. The good agreement between the measured and calculated interaction energies allows us to use the results from this calculation as the calibration for $U$ for all our measurements.

\section{Preparation of eigenstates of the symmetric double well}
\label{sec:pass_eigenstates}

\begin{figure*}
	\centering
	\includegraphics[width=0.9\textwidth]{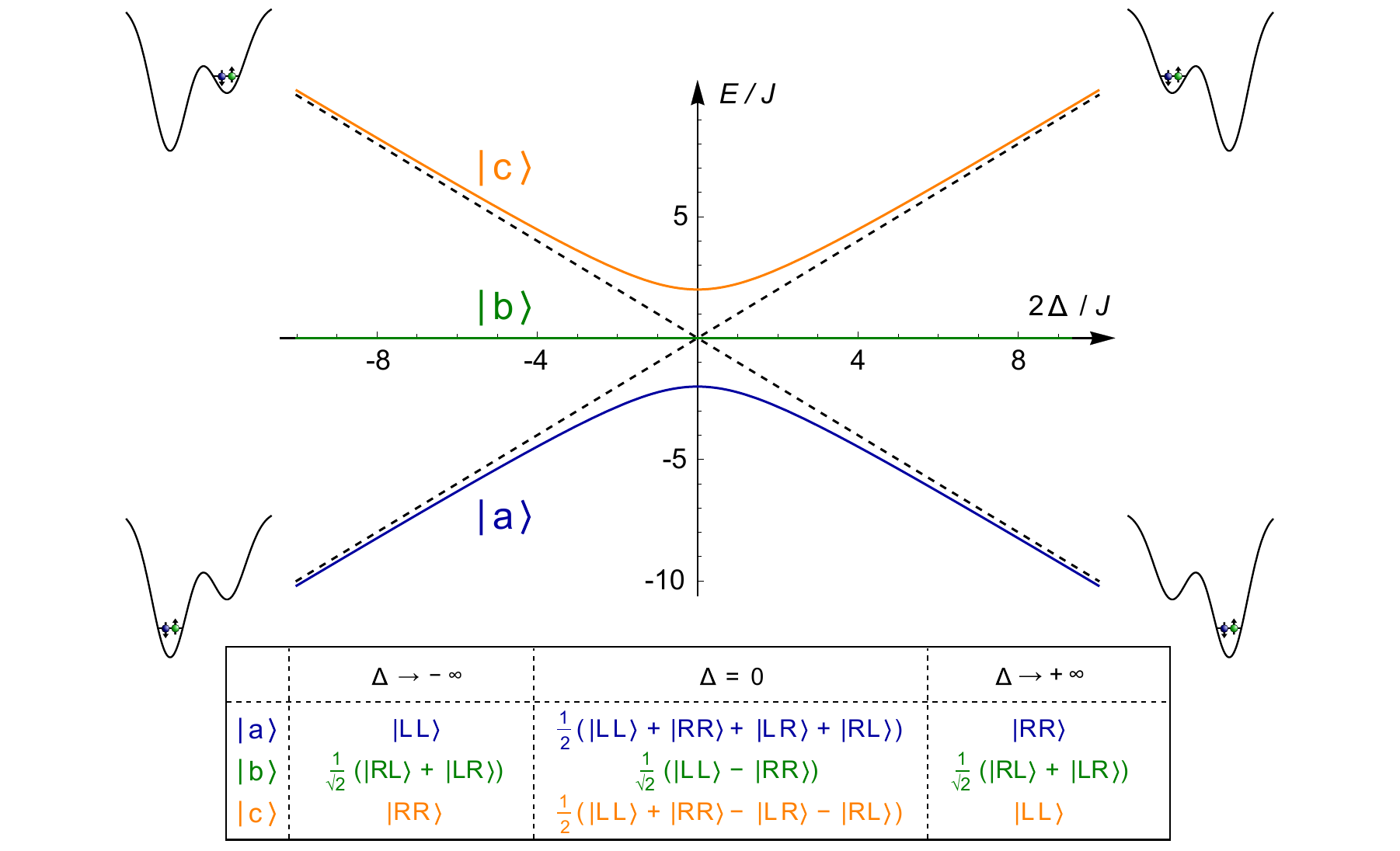}
	\caption{Eigenenergies and eigenstates in a tilted double well. Energies of states $|a\rangle$, $|b\rangle$, and $|c\rangle$ in a non-interacting system as a function of the tilt $\Delta$ between the wells. Depending on the sign of the initial tilt, state $|LL\rangle$ can be used to prepare either state $|a\rangle$ or state $|c\rangle$.}
	\label{fig:som_gs_preparation}
\end{figure*}

To prepare two atoms in an eigenstate of the symmetric double well, we start with two non-interacting atoms in the ground state of a single well and adiabatically change the shape of the potential to a double well. To prepare the system in the ground state $|a\rangle$, we initialize the system in state $|LL\rangle$, but keep the left well much deeper than the right one. Then, we lower the potential barrier between the wells and thereby set the tunnel coupling $J$ to a value of about $h\times\unit[100]{Hz}$. By setting a tilt of $|\Delta| \approx \frac{1}{2} \hbar \omega_{ax}$, we make sure that tunneling is far off-resonant ($|\Delta| \gg J$) and the initial state $|LL\rangle$ has almost complete overlap with the ground state of the tilted double well (Fig.\,4). By adiabatically reducing the tilt to $\Delta = 0$ within 100\,ms we then bring the system into the ground state of the symmetric double well.

Using the same technique, we can also prepare the atoms in the highest energy state $|c\rangle$. To do this, we start again in state $|LL\rangle$, but this time we set the right well to be much deeper than the left one ($\Delta \approx + \frac{1}{2} \hbar \omega_{ax}$). Lowering the barrier then transforms the initial state $|LL\rangle$ into state $|c\rangle$ (Fig.\,4).

After preparing the non-interacting eigenstate of the double-well system, we adiabatically introduce interparticle interactions by ramping the magnetic field from the zero-crossing of the scattering length at 527\,G to values between $\unit[300]{G}$ and $\unit[740]{G}$ over 60\,ms. This corresponds to scattering lengths between $\unit[-288]{a_0}$ and $\unit[2974]{a_0}$ and interaction energies between $U/h\approx -1.3\,J/h \approx  -87\,$Hz and $U/h\approx10.1\,J/h \approx 680$\,Hz, where $a_0$ is the Bohr radius (Fig.\,3).

Similar to the experimental sequence presented here, one can also start from an interacting system with large tilt and both particles in the deeper well and then adiabatically go to the ground state of the symmetric double well \cite{Petta2005}.

To confirm the adiabaticity of the ramps in tilt and in interaction strength, we drive the system from state $|LL\rangle$ to the final configuration, hold the system for different durations and then reverse both ramps. To check for excitations, we measure the mean atom number in the ground state both with and without the ramps. Within the statistical uncertainty of our measurement we could not detect any heating caused by the ramps. From these measurements, we determine a fidelity of $91.6\pm0.7$\% of preparing the system in the ground state of the symmetric double well.

\section{Number statistics and preparation fidelity correction}

\begin{figure*}
	\centering
	\includegraphics[width=0.5\textwidth]{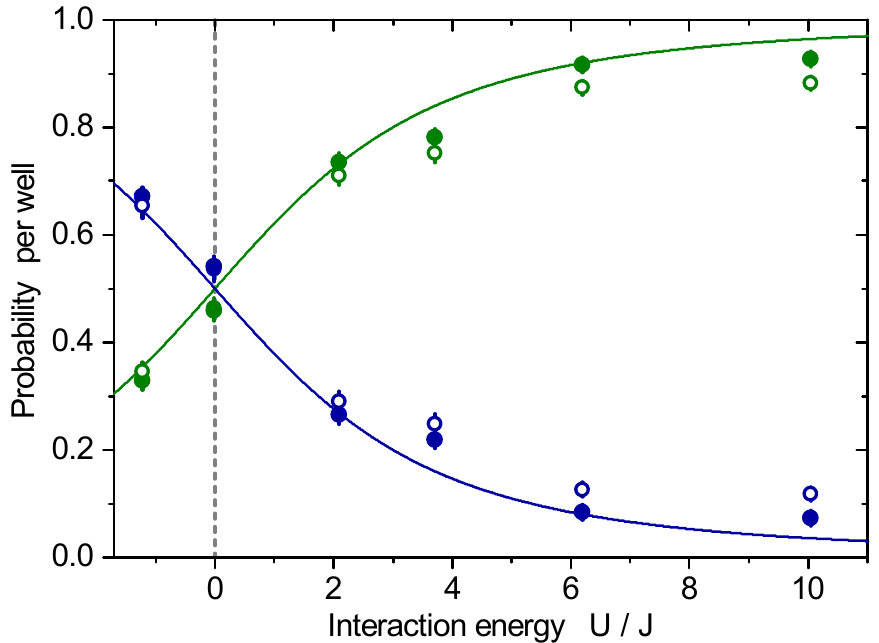}
	\caption{Corrected and non-corrected occupation statistics. Single occupancy (green) and double occupancy (blue) in state $|a\rangle$ as a function of the on-site interaction energy $U$ at a tunnel coupling of \mbox{$J/h \simeq \unit[67]{Hz}$}. Open symbols correspond to the occupation statistics which were directly calculated from the measured probabilities of finding 0, 1, or 2 atoms in one of the wells. For the filled symbols, these probabilities were corrected for the finite experimental fidelities in the preparation and detection processes before calculating the occupation statistics. The solid lines show the prediction of the Hubbard model. The error bars denote the $1\sigma$ statistical uncertainties.}
	\label{fig:som_stat_graph}
\end{figure*}

To observe the transition from the metallic to the insulating regime, we measure the probability of single and double occupancy in the symmetric double well as a function of interaction strength (Fig.\,3 of main text). Hence, we first prepare the system either in state $|a\rangle$ or state $|c\rangle$ at different interaction strengths (see Sec.\,\ref{sec:pass_eigenstates}). Then, we quickly ramp up the potential barrier to decouple the wells and measure the atom number in either the left or the right well. We repeat this measurement 300 times for each well and count how often we detect zero atoms ($N_0$), one atom $(N_1$), or two atoms ($N_2$). The probabilities of measuring $0$, $1$, or $2$ atoms follow as: $a_i' = N_i / (N)$, where $i \in {0, 1, 2}$ and $N$ is the total number of measurements in both wells.

In our measurement, we have finite fidelities both for the preparation of the initial state and the detection of the final atom number. Both the preparation fidelity and the detection fidelity reduce the measured atom numbers as compared to an ideal measurement. In a symmetric double well the influence of these processes on the measured probabilities cannot be distinguished. We therefore combine the preparation and detection fidelity into an overall fidelity $p$ which we define as the mean atom number per well. We determine $p$ by averaging over all individual measurements. 

Due to the finite fidelity $p < 1$ of our experiment, the measured probabilities $a_0'$, $a_1'$, and $a_2'$ deviate from the probabilities $a_0$, $a_1$, and $a_2$ of an ideal experiment with $p = 1$:
\begin{align}
	a_2'& = a_2 p^2 \\
	a_1'& = a_1 p + 2 a_2 p (1-p) \\
	a_0'& = a_0 + a_1 (1-p) + a_2 (1-p)^2
\end{align}
These three equations can be inverted to calculate the ideal probabilities:
\begin{align}
	a_2& = a_2'/p^2 \\
	a_1& = a_1'/p - 2 (1-p) a_2'/p^2 \\
	a_0& = a_0' - a_1' (1-p)/p + a_2' (1-p)^2 / p^2
\end{align}
Using these corrected probabilities, we define single occupancy as the probability of measuring a single atom ($P_1 = a_1$) and double occupancy as the sum of the probabilities of measuring either zero or two atoms ($P_2 = a_0+a_2$). For comparison, the results for single occupancy and double occupancy in the ground state $|a\rangle$ using both the corrected and the non-corrected probabilities are shown in Fig.\,5. To verify that our measurement of the occupation statistics is insensitive to small drifts of the tilt, we performed the measurement for state $|c\rangle$ at different tilts in a range of $\pm \unit[30]{Hz}$ around $\Delta=0$. We observe that while the tilt affects the probabilities $a_0$ and $a_2$ of finding zero or two particles in the individual wells there is no significant change in the double occupancy $a_0+a_2$.

\section{Trap modulation spectroscopy}

\begin{figure*}
	\centering
	\includegraphics[width=\textwidth]{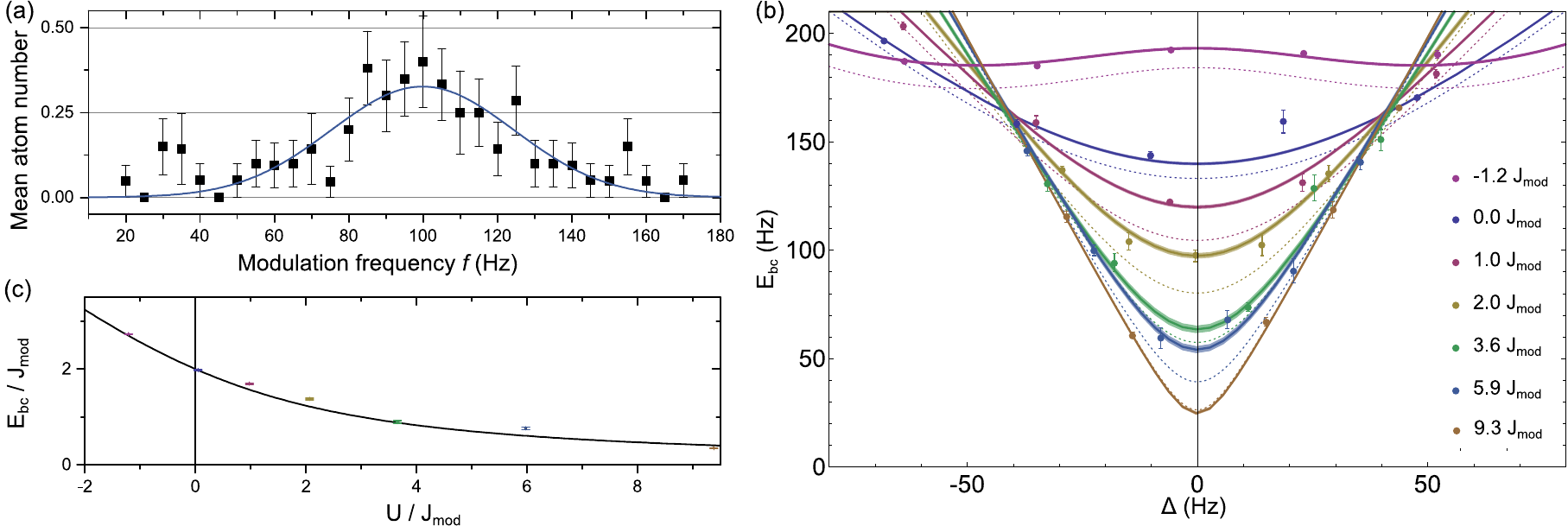}
	\caption{Energy difference $E_{bc}$ between states $|c\rangle$ and $|b\rangle$. (a) Mean atom number in the right well at different modulation frequencies for \mbox{$U=5.9 J_\text{mod}$} and \mbox{$\Delta/h = \unit[-22]{Hz}$}. The resonance frequency is extracted from a Gaussian fit to the peak (blue curve). ( b) Resonance frequency of the transition between states $|c\rangle$ and $|b\rangle$ as a function of the tilt $\Delta$ for different interaction energies. The predictions by the Hubbard model without free parameters are shown as dashed lines. The solid curves are fits to the data using the Hubbard model with $J$ as a free parameter, the shaded areas are the respective $1\sigma$ confidence bands for the mean predictions. (c) Energy difference $E_{bc}$ of the balanced double well plotted against the interaction energy $U$. The errors correspond to the width of the respective confidence bands at $\Delta = 0$ in the right panel.}
	\label{fig:som_dw_spectroscopy}
\end{figure*}

To directly observe the effect of the tunnel coupling on the energy of our double-well system, we measure the energy difference $E_{bc} = E_c-E_b$ between state $|c\rangle$ and state $|b\rangle$ by trap modulation spectroscopy. We start by initializing state $|c\rangle$ with a tunnel coupling of $J=h\times\unit[67.3]{Hz}$ and interaction energies between $U=-1.3\,J$ and $U=10.1\,J$, as described above. Then we perform a sinusoidal modulation of the total trapping power with frequencies between $f = \unit[30]{Hz}$ and $f = \unit[300]{Hz}$ for a duration of $\unit[200]{ms}$. The atoms are transferred resonantly to state $|b\rangle$ if the modulation frequency matches the energy difference $E_{bc}$. To minimize the distortion of the energy spectrum, we modulate the tunnel coupling only with a small amplitude of $0.11\,J$. To determine whether the atoms have been transferred to state $|b\rangle$, we adiabatically turn off the interactions, ramp to a large tilt $\Delta \gg 0$ and measure the number of atoms in the right well. If the system has remained in state $|c\rangle$ during the modulation, both atoms are then in the left well. However, if the system has been transferred to state $|b\rangle$, there is a large probability to observe a single atom in the right well for $\Delta \gg 0$ (Fig.\,4).  We plot the mean atom number in the right well as a function of the modulation frequency and extract the resonance frequency with a Gaussian fit (Fig.\,6a).

To measure the effects of second-order tunneling, we have to control the tilt better than $4J^2 /U$, which is approximately $\unit[30]{Hz}$ at $U\approx10J$. This is comparable to the residual drifts in our calibration of the tilt $\Delta$ over the course of several days (see Sec.\,\ref{sec:tilt}). We solve this problem by measuring the resonance frequency for different tilts around $\Delta = 0$ for each of the interaction energies. We determine the shift $\Delta_0 = \Delta' - \Delta$ between our set value $\Delta'$ and the tilt $\Delta$ in the experiment for each data set by fitting a Rabi function $f_\text R = 2 \sqrt{J^2 + (\Delta'-\Delta_0)^2}$ with $\Delta_0$ and $J$ as free parameters. We then center our data by subtracting the respective $\Delta_0$.

For a non-interacting system, the energy difference $E_{bc}$ is given by the transition frequency $f_\text R(\Delta = 0)$. For an interacting system, we can still use this fit to center our data, but it does not yield the correct value for $E_{bc}$. Hence, we compare our measurements to a numeric solution of the Hubbard model with fixed parameters $U$ and $J$ as calibrated from the tunneling measurements (dashed curves in Fig.\,6b). We observe resonance frequencies which are systematically higher than predicted. Therefore, we fit the data with a numeric Hubbard model with $J'$ as a free parameter (solid lines). We obtain $E_{bc}(\Delta = 0)$ for the balanced double-well system at different values of $U$ from the fitted curves (Fig.\,6c and Fig.\,4 of the main text).

One possible cause for the deviation of the fixed-parameter Hubbard model from our data is that the modulation of the trapping potential results in a modified tunnel coupling. Therefore, we scale $E_{bc}$ and $U$ with the modified tunnel coupling $J_\text{mod} = h\times\unit[(70.7 \pm 0.3)]{Hz}$ which we obtain from the fit at $U=0$.

\end{document}